\documentclass[12pt,a4paper]{article}
\usepackage{graphicx}
\usepackage{amsmath}
\usepackage{amsthm}
\numberwithin{equation}{section}
\advance\textheight4cm
\advance\topmargin-2cm
\advance\oddsidemargin-0.5cm
\advance\evensidemargin-0.5cm
\advance\textwidth1cm
\numberwithin{equation}{section}
\title{On systems of interacting populations influenced by  
multiplicative white noise}
\author{Nikolay K. Vitanov, Kaloyan N. Vitanov}
\date{Institute of Mechanics, Bulgarian Academy of Sciences \\
Acad. G. Bonchev Str., Bl. 4, 1113 Sofia, Bulgaria}
\begin{document}
\maketitle
\begin{abstract}
We discuss a model of a system of interacting populations for the case
when: (i) the growth rates and the coefficients of interaction among the
populations depend on the populations densities: and (ii) the environment 
influences the growth rates and this influence can be modelled by a Gaussian
white noise. The system of model equations for this case is a system of
stochastic differential equations with: (i) deterministic part in the form of
polynomial nonlinearities; and (ii) state-dependent stochastic part in the
form of multiplicative Gaussian white noise. We discuss both the cases when
the formal integration of the stochastic differential equations leads: (i)
to integrals of Ito kind; or (ii) to integrals of Stratonovich kind. The
systems of stochastic differential equations are reduced to the corresponding
Fokker-Planck equations. For the Ito case and for the case of 1 population am
analytic results is obtained for the stationary PDF of the the population
density. For the case of  more than one population and for the both Ito case 
and Stratonovich case the detailed balance conditions are not satisfied and
because of this exact analytic solutions of the corresponding Fokker-Plank
equations for the stationary PDFs for the population densities are not
known. We obtain approximate solutions for this case by the
methodology of the adiabatic elimination. 
\end{abstract}
\section{Introduction}
The research on the nonlinear dynamics of the complex systems increases
steadily in the last two decades (for several examples see Appendix A). Many
complex systems are influenced by random events. Because of this the theory
of stochastic processes is much used in the modeling of the processes in the
complex systems \cite{st1}-\cite{st5}.
In this paper we discuss some mathematical aspects of the theory of interacting
populations for the case when the growth rates are influenced by
environmental fluctuations. For the case when the fluctuations can be modelled
by Gaussian white noise we shall obtain as model equations a system
of stochastic differential equations that contain multiplicative
noise. For the case of single population the model equation will be
of the kind
\begin{equation}\label{a1}
{\dot{\rho}} = F(\rho) + \eta G(\rho)
\end{equation}
where $F(\rho)$ and $G(\rho)$ are polynomials of $\rho$ and $\eta$ is
Gaussian white noise. For the case of system of interacting populations the
corresponding model equations will be of the kind
\begin{equation}\label{a2}
{\dot{\rho}_i} = F_i(\rho_1,\dots,\rho_n) + \eta_i
 G_i(\rho_1,\dots,\rho_n); \ i=1,\dots,n
\end{equation}
where $F$ and $G$ are polynomials and $\eta_i$ are Gaussian white noises.
\par
The organization of the article is as follows. We discuss the model equations
for the dynamics of interacting populations in the following section. The
presence of Gaussian white noise in the growth rates of populations leads
to a system of stochastic differential equations with multiplicative noise.
The integration of these stochastic differential equations leads in
principle to stochastic integrals of Ito kind or to stochastic integrals of
Stratonovich kind. Section 3 is devoted to the theory for the case when the
stochastic integrals are of Ito kind. Section 4 is devoted to theory for the
case when the stochastic integrals are of Stratonovich kind. Several concluding
remarks are summarized in Section 5. In addition four appendices supply
the reader with information about the examples of research on complex systems,
about the theory of  stochastic differential
equations containing multiplicative white noise, theory of stochastic
differential equations of Ito and Stratonovich kind and their relation to the
Fokker-Plank equation (known also as forward Kolmogorov equation).
\section{Investigated equations and population dynamics}
The classical model of interacting populations is based on a system of
equations of Lotka-Volterra kind \cite{murr,turchin}:
\begin{equation}\label{b1}
\dot{\rho}_i = r_i \rho_i(t) \left( 1- \sum_{j=1}^n \alpha_{ij} \rho_j(t) \right)
\end{equation}
where $\rho_i$ are the densities of the population members, $r_i$ are the 
birth rates, and $\alpha_{ij}$ are coefficients of
interaction between the populations $i$ and $j$.
Let us now suppose that the birth rates and interaction coefficients
depend on the density of the populations and in addition the birth rates
fluctuate:
\begin{equation}\label{b2}
r_i = r_i^0 \left( 1 + \sum_{j=1}^n r_{ij} \rho_j\right) + \eta_i; \ \
\alpha_{ij} = \alpha_{ij}^0 \left(1+ \sum_{j=1}^n \alpha_{ijk} \rho_k \right)
\end{equation} 
in Eq. (\ref{b2}) $r_{ij}$ and $\alpha_{ijk}$ are parameters and $\eta_i$
are Gaussian white noises. The system of equations (\ref{b2}) for $\eta_i=0$ 
has been introduced and investigated in \cite{dv1}-\cite{vjd2}. The presence
of $\eta_i$ however influences much the system dynamics \cite{bm1,vdv12}.
\par
The substitution of Eq.(\ref{b2}) in Eq.(\ref{b1}) leads to a system
of model equations of the kind
\begin{eqnarray}\label{b3}
{\dot{\rho}_i} &=& F_i(\rho_1,\dots,\rho_n) + \eta_i G_i(\rho_1,\dots,\rho_n);
\nonumber \\
 F_i(\rho_1,\dots,\rho_n) &=& r_i^0 \rho_i \bigg \{ 1 -\sum_{j=1}^n
 (\alpha_{ij}^0 - r_{ij}) \rho_j - \sum_{j=1}^n 
 \sum_{l=1}^n \alpha_{ij}^0 (\alpha_{ijl} + r_{il}) \rho_j \rho_l -
 \nonumber \\
 && \sum_{j=1}^n \sum_{k=1}^n \sum_{l=1}^n \alpha_{ij}^0 r_{ik} \alpha_{ijl}
 \rho_j \rho_k \rho_l \bigg \}  \nonumber \\
 G_i(\rho_1,\dots,\rho_n) &=& \rho_i \left(1 - \sum_{j=1}^n \alpha_{ij}^0 \rho_j
 - \sum_{j=1}^n \sum_{k=1}^n \alpha_{ij}^0 \alpha_{ijk} \rho_j \rho_k \right) 
\end{eqnarray}
\par
For the case of one population (we set $r^0=r$; $r_{11}=0$; $\alpha_{11}^0
=\alpha$; $\alpha_{111}=0$) the model equation is
\begin{eqnarray}\label{b4}
\dot{\rho} &=& F(\rho) + \eta G(\rho) \nonumber \\
F(\rho)&=& r \rho - \alpha r \rho^2; \ \ G(\rho) = \rho - \alpha \rho^2
\end{eqnarray}
Below we shall discuss more general equation in comparison to Eq.(\ref{b4}). 
We shall discuss the case where $F(\rho)$ and $G(\rho)$
are polynomials of arbitrary orders $p_1$ and $p_2$, i.e.,
\begin{equation}\label{b4a}
F(\rho)= \sum_{i=1}^{p_1} \mu_i \rho^i; \ \ G(\rho) = \sum_{i=1}^{p_2} 
\theta_i \rho^i
\end{equation}
where $\mu_i$ and $\theta_i$ are parameters.
In this case Eq.(\ref{b4}) becomes
\begin{eqnarray}\label{b4b}
\dot{\rho} &=& \sum_{i=1}^{p_1} \mu_i \rho^i + 
\eta \sum_{i=1}^{p_2} 
\theta_i \rho^i \nonumber \\
\end{eqnarray}
The formal integration of Eq.(\ref{b4}) (see also Appendix B) leads to the 
equation
\begin{equation}\label{b5}
\rho (t) = \rho(t=0) + \int_0^t d \tau \ F[\rho(\tau)] + 
\int_0^t dW_\tau \ \ G[\rho(\tau)],
\end{equation}
where $W_\tau$ is a Wiener process.
The integral $\int\limits_0^t dW_\tau \ \ G(\rho(\tau))$ can be integral of Ito kind
or integral of Stratonovich kind (for more discussion see  Appendix
B). In the next two sections we shall discuss these two cases.  
\section{Case of stochastic differential equations of Ito kind}
For this case Eq.(\ref{b4b}) can be written as
\begin{equation}\label{c1}
d\rho_t =  F(\rho_t)dt +  G(\rho_t) dW_t,
\end{equation}
where we denote the time dependence as subscript and in general $F$ and $G$
are given by Eqs.(\ref{b4a}). The Fokker-Planck equation that corresponds to 
Eq.(\ref{c1}) is
\begin{eqnarray}\label{c2}
\frac{\partial }{\partial t}p(x,t) = - \frac{\partial}{\partial x} \Bigg \{
p(x,t) \bigg[ \sum_{i=1}^{p_1} \mu_i x^i \bigg] \Bigg \} +
\frac{1}{2} \frac{\partial^2}{\partial x^2} \Bigg \{p(x,t) \bigg[ 
\sum_{i=1}^{p_2} \sum_{j=1}^{p_2} \theta_i \theta_j x^{i+j} \bigg] \Bigg \}
\nonumber \\
\end{eqnarray}
We can formulate the following
\newtheorem*{prop1}{Proposition 1}
\begin{prop1}
Let $b_1$ and $b_2$ be natural boundary points ($-\infty \le b_1 < b_2 \le 
\infty $). Let in addition $\sigma(x) = \sum\limits_{i=1}^{p_2} \theta_i x^i >0$ in $(b_1,b_2)$. Then
the diffusion process $X_t$ that is solution of the stochastic differential
equation Eq.(\ref{c1}) has unique invariant distribution with
p.d.f.
\begin{equation}\label{c3}
p^0(x) = \frac{\cal{N}}{\sum\limits_{i=1}^{p_2} \sum\limits_{j=1}^{p_2} 
\theta_i \theta_j x^{i+j}} \exp \left(  \int_c^x dy 
\frac{2 \sum\limits_{i=1}^{p_1} \mu_i y^i }{\sum\limits_{i=1}^{p_2}
\sum\limits_{j=1}^{p_2} \theta_i \theta_j y^{i+j}} \right), \ \ \vee x \in (b_1,b_2)
\end{equation}
if the quantity
\begin{equation}\label{c4}
{\cal{N}}^{-1}= \int_{b_1}^{b_2} dx \frac{1}{\sum\limits_{i=1}^{p_2} 
\sum\limits_{j=1}^{p_2} \theta_i \theta_j x^{i+j}} \exp \left( \int_c^x 
dy \frac{2 \sum\limits_{i=1}^{p_1} \mu_i y^i}{\sum\limits_{i=1}^{p_2}
\sum\limits_{j=1}^{p_2} \theta_i \theta_j y^{i+j}} \right), \ \ b_1 < c < b_2
\end{equation}
has finite value. In addition each time-dependent solution $p(x,t)$ of the
Fokker-Planck equation (\ref{c2}) in $(b_1,b_2)$ satisfies
\begin{equation}\label{c5}
\lim_{t \to \infty} p(x,t) = p^0(x)
\end{equation}
\end{prop1}
\begin{proof}
The proposition follows from the Observation 1 from the Appendix B
for the case when
\begin{equation*}
f(x)= \sum_{i=1}^{p_1} \mu_i x^i; \ \ \sigma(x) = \sum_{i=1}^{p_2} 
\theta_i x^i.
\end{equation*}
\end{proof}
\par
Let us apply the Proposition 1 to the case of one population modelled by
Eq.(\ref{b4}). In this case $\mu_1=r$; $\mu_2=-\alpha r$; $\theta_1=1$; 
$\theta_2=-\alpha$. We note that Proposition 1 is valid when $\sigma>0$.
In our case this means that $\rho < 1/\alpha$ ($\rho \ge 0$). For the
quantity $N$ from Eq.(\ref{c4}) we obtain 
\begin{eqnarray}\label{c6}
{\cal{N}}^{-1} = \frac{(\alpha c -1)^{2r}}{r(4r^2-1)c^{2r}} \bigg[ 
\frac{b_2^{2r-1} ((r-\alpha b_2)(2r+1) + \alpha^2 b_2^2 )}{(1 - \alpha
b_2)^{2r+1}} - \nonumber \\
\frac{b_1^{2r-1} ((r-\alpha b_1)(2r+1) + \alpha^2 b_1^2 )}{(1 - \alpha
b_1)^{2r+1}}
\bigg]
\end{eqnarray}
and for $p^0(x)$ from Eq.(\ref{c3}) we obtain
\begin{eqnarray}\label{c7}
p^0(x) =r(1-4r^2) \bigg[ - 
\frac{b_2^{2r-1} ((r-\alpha b_2)(2r+1) + \alpha^2 b_2^2 )}{(1 - \alpha
b_2)^{2r+1}} + \nonumber \\
\frac{b_1^{2r-1} ((r-\alpha b_1)(2r+1) + \alpha^2 b_1^2 )}{(1 - \alpha
b_1)^{2r+1}}
\bigg]^{-1} \frac{1}{x^{2-2r}(1 - \alpha x)^{2r+2}}
\end{eqnarray}
Let us now discuss the case of more than one population. For this case
we have to solve the system of stochastic differential equations
\begin{equation}\label{c8}
dX_i(t) = F_i[X_1(t),\dots,X_n(t)] +  G_{i}[X_1(t),
\dots,X_n(t)] dW_i(t), \ i=1,\dots,n
\end{equation}
where $W_j(t)$ are independent Wiener processes and 
\begin{eqnarray}\label{c9}
 F_i(\rho_1,\dots,\rho_n) &=& r_i^0 \rho_i \bigg \{ 1 -\sum_{j=1}^n
 (\alpha_{ij}^0 - r_j) \rho_j - \sum_{j=1}^n 
 \sum_{l=1}^n \alpha_{ij}^0 (\alpha_{ijl} + r_{il}) \rho_j \rho_l -
 \nonumber \\
 && \sum_{j=1}^n \sum_{k=1}^n \sum_{l=1}^n \alpha_{ij}^0 r_{ik} \alpha_{ijl}
 \rho_j \rho_k \rho_l \bigg \}  \nonumber \\
 G_i(\rho_1,\dots,\rho_n) &=& \rho_i \left(1 - \sum_{j=1}^n \alpha_{ij}^0 \rho_j
 - \sum_{j=1}^n \sum_{k=1}^n \alpha_{ij}^0 \alpha_{ijk} \rho_j \rho_k \right) 
\end{eqnarray}
The corresponding Fokker-Planck equation is ($G_{ij}=G_i \delta_{ij}$ where
$\delta_{ij}$ is the Kronecker delta-symbol)
\begin{eqnarray}\label{c10}
\frac{\partial}{\partial t} p = - \sum_{i=1} \frac{\partial}{\partial x_i}
[p F_i (x_1,\dots,x_n,t)] + \nonumber \\ \frac{1}{2} \sum_{i=1}^n \sum_{j=1}^m
\frac{\partial}{\partial x_i} \frac{\partial}{\partial x_j}[p
G_{ij}(x_1,\dots,x_n,t)G_{ji}(x_1,\dots,x_n,t)]
\end{eqnarray}
We are interested in stationary solutions $p_s$ of Eq.(\ref{c10}). In the general
case such solutions can be obtained numerically. A hope to obtain analytic
solutions exists mainly  when the conditions for detailed balance are satisfied
\cite{gard1}. For the case of Eq.(\ref{c10}) these conditions are
\begin{eqnarray}\label{c11}
\epsilon_i F_i (\vec{\epsilon} \cdot \vec{x}) p_s (\vec{x}) &=& - 
F_i(\vec{x}) p_s(\vec{x}) + \sum_{j=1}^n \frac{\partial}{\partial x_j}
[G_i^2(\vec{x}) p_s(\vec{x})] \nonumber \\
\epsilon_i^2 G_i^2(\vec{\epsilon} \cdot \vec{x}) &=& G_i^2(\vec{x})
\end{eqnarray}
where $\epsilon_i = \pm 1$. One can easily show that that the structure
of $G_i$ from (\ref{c9}) is such that the second condition for existence
of detailed balance from (\ref{c11}) is not satisfied. Then
one can hope to obtain approximate analytic solutions for particular
cases of Eq.(\ref{c10}).
\par
One possible way for obtaining approximate solutions of the Fokker-Planck
equation for the case of more than one population is the connected to the
method of adiabatic elimination \cite{gard1}. In order to illustrate
this method we consider the following particular case of the Eqs.(\ref{b3}).
Let $\alpha_{ij}^0 = 0$ and $\alpha_{ijk}=0$. In addition let $\eta_2 =0$.
For the case of two populations we obtain the following system of equations
\begin{eqnarray}\label{c12}
d{\rho}_1 &=& r_1^0 \rho_1 (1 + r_{11} \rho_1 + r_{12} \rho_2) dt + \rho_1 dW_1
\nonumber \\
d{\rho}_2 &=& r_2^0 \rho_2 (1+ r_{21} \rho_1 + r_{22} \rho_2 )dt
\end{eqnarray}
Let $\rho_2$ be the fast relaxing variable,i.e., $d \rho/dt$ tends to $0$
very fast in the time. Then in the second equation of Eqs.(\ref{c12})
one can set $d \rho /dt =0$ and then the resulting equation has solutions
$\rho_2 =0$ (extinction of the second population) or
\begin{equation}\label{c13}
\rho_2 = - \frac{1}{r_{22}} - \frac{r_{21}}{r_{22}}\rho_1
\end{equation}
which corresponds to a "slaving" of the "fast" variable $\rho_2$
by the "slow" variable $\rho_1$. The substitution of Eq.(\ref{c13})
in the first equation of Eqs.(\ref{c12}) leads to the stochastic differential
equation
\begin{equation}\label{c14}
d \rho_1 = \left[ r_1^0 \left( r_{11} - \frac{r_{12} r_{21}}{r_{22}} \right) 
\rho_1^2 +
r_1^0 \left( 1 - \frac{r_{12}}{r_{22}}\right) \rho_1 \right] dt + \rho_1 dW_1
\end{equation}
Eq.(\ref{c14}) can be treated by the methodology discussed above.
In order to obtain an analytic result we have to assume
\begin{equation}\label{c15}
r_1^0 = \frac{2}{1-r_{12}/r_{22}}
\end{equation}
The application of the methodology connected to \textit{Proposition 1}
leads to the distribution
\begin{equation}\label{c16}
p^0(\rho_1) = A \rho_1^2 \exp(2 \mu_2 \rho_1) 
\end{equation}
where
\begin{equation*}
A = \frac{\mu_2^3}{\left( \cfrac{\mu_2^2 b_2^2}{2}-\cfrac{\mu_2 b_2}{2}+
\cfrac{1}{4} \right) \exp(2 \mu_2 b_2) - \left( \cfrac{\mu_2^2 b_1^2}{2} -
\cfrac{\mu_2 b_1}{2} + \cfrac{1}{4} \right) \exp(2 \mu_2 b_1)} 
\end{equation*}
and
\begin{equation}\label{c17}
\mu_2 = \frac{2 r_{11} (r_{22}-r_{11})}{r_{22}-r_{12}} <0
\end{equation}
\section{Case of stochastic differential equations of Stratonovich kind}
For this case Eq.(\ref{b4b}) can be written as
\begin{equation}\label{d1}
d\rho_t =  \left[ F(\rho_t)+ \frac{1}{2} G'(\rho_t) G(\rho_t) \right] dt +  
G(\rho_t) dW_t,
\end{equation}
where we again again denote the time dependence as subscript and in general $F$ and $G$
are given by Eqs.(\ref{b4a}). According to Appendix B the Fokker-Planck equation
that corresponds to Eq.(\ref{d1})
is
\begin{eqnarray}\label{d2}
\frac{\partial }{\partial t}p(x,t) = - \frac{\partial}{\partial x} \Bigg \{
p(x,t) \bigg[ \bigg( \sum_{i=1}^{p_1} \mu_i \rho^i \bigg) + \frac{1}{2}
\bigg( \sum_{i=1}^{p_1} \sum_{j=1}^{p_2} i \theta_i \theta_j \rho^{i+j-1}\bigg)
\bigg] \Bigg \} + \nonumber \\
\frac{1}{2} \frac{\partial^2}{\partial x^2} \Bigg \{p(x,t) \bigg[ 
\sum_{i=1}^{p_2} \sum_{j=1}^{p_2} \theta_i \theta_j \rho^{i+j} \bigg] \Bigg \}
\nonumber \\
\end{eqnarray}
We can formulate the following
\newtheorem*{prop2}{Proposition 2}
\begin{prop2}
Let $b_1$ and $b_2$ be natural boundary points ($-\infty \le b_1 < b_2 \le 
\infty $). Let in addition $\sigma(x)= \sum\limits_{i=1}^{p_2} \theta_i x^i  >0$ in $(b_1,b_2)$. Then
the diffusion process $X_t$ that is solution of the stochastic differential
equation Eq.(\ref{d1}) has unique invariant distribution with
p.d.f.
\begin{eqnarray}\label{d3}
p^0(x) = \frac{\cal{N}}{\sum\limits_{i=1}^{p_2} \sum\limits_{j=1}^{p_2} 
\theta_i \theta_j x^{i+j}} \exp \left(  \int_c^x dy 
\frac{2 \left( \sum\limits_{i=1}^{p_1} \mu_i y^i  + \frac{1}{2}
\sum\limits_{i=1}^{p_1} \sum\limits_{j=1}^{p_2} i \theta_i \theta_j y^{i+j-1}\right) }{\sum\limits_{i=1}^{p_2}
\sum\limits_{j=1}^{p_2} \theta_i \theta_j y^{i+j}} \right), \nonumber \\
 \vee x \in (b_1,b_2), \nonumber \\
\end{eqnarray}
if the quantity
\begin{eqnarray}\label{d4}
{\cal{N}}^{-1}= \int_{b_1}^{b_2} dx \frac{1}{\sum\limits_{i=1}^{p_2} 
\sum\limits_{j=1}^{p_2} \theta_i \theta_j x^{i+j}} \exp \left( \int_c^x 
dy \frac{2 \left( \sum\limits_{i=1}^{p_1} \mu_i y^i + \frac{1}{2}
\sum\limits_{i=1}^{p_1} \sum\limits_{j=1}^{p_2} i \theta_i \theta_j y^{i+j-1}\right)}{\sum\limits_{i=1}^{p_2}
\sum\limits_{j=1}^{p_2} \theta_i \theta_j y^{i+j}} \right), \nonumber \\
 b_1 < c < b_2, \nonumber \\
\end{eqnarray}
has finite value. In addition each time-dependent solution $p(x,t)$ of the
Fokker-Planck equation (\ref{d2}) in $(b_1,b_2)$ satisfies
\begin{equation}\label{d5}
\lim_{t \to \infty} p(x,t) = p^0(x)
\end{equation}
\end{prop2}
\begin{proof}
The proposition follows from the Observation 2 from the Appendix C
for the case when
\begin{equation*}
f(x)= \sum_{i=1}^{p_1} \mu_i x^i; \ \ \sigma(x) = \sum_{i=1}^{p_2} 
\theta_i x^i.
\end{equation*}
\end{proof}
\par
Let us apply the Proposition 2 to the case of one population modelled by
Eq.(\ref{b4}). In this case $\mu_1=r$; $\mu_2=-\alpha r$; $\theta_1=1$; 
$\theta_2=-\alpha$. We note that Proposition 2 is valid when $\sigma>0$.
In our case this means that $\rho < 1/\alpha$ ($\rho \ge 0$). For the
quantity $N$ from Eq.(\ref{d4}) we obtain
\begin{equation}\label{d6}
{\cal{N}}^{-1} = \int_{b_1}^{b_2} dx \ \frac{(\alpha c - 1)^{2r-3} x^{2r-1}}{
c^{2r+1}(\alpha x - 1)^{2r-1}}
\exp \bigg[-2 \alpha c + \cfrac{2 \alpha x}{(\alpha x - 1)(\alpha c - 1)} \bigg]
\end{equation}
and for the distribution $p^0(x)$ we obtain
\begin{eqnarray}\label{d7}
p^0(x) = \cfrac{\bigg[  \cfrac{1}{x} \bigg( \cfrac{1}{\alpha x - 1}\bigg)^{2r-1} \bigg]
\exp \bigg[-2 \alpha c + \cfrac{2 \alpha x}{(\alpha x - 1)(\alpha c - 1)} \bigg]}{
\int_{b_1}^{b_2} dy \ \cfrac{ y^{2r-1}}{(\alpha y - 1)^{2r-1}}
\exp \bigg[-2 \alpha c + \cfrac{2 \alpha y}{(\alpha y - 1)(\alpha c - 1)} \bigg]}
\nonumber \\
\end{eqnarray}
\par
Let us now discuss the case of more than one population. For this case
we have to solve the system of stochastic differential equations
\begin{eqnarray}\label{d8}
dX_i(t) = \bigg \{ F_i[X_1(t),\dots,X_n(t)] + \nonumber \\
\frac{1}{2} G_{i}(X_1(t),\dots,X_n(t))\frac{\partial }{\partial
x_i}[G_{i}(X_1(t),\dots,X_n(t))] \bigg \} 
+ \nonumber \\ G_{i}[X_1(t),
\dots,X_n(t)] dW_i(t), \ i=1,\dots,n
\end{eqnarray}
where $W_j(t)$ are independent Wiener processes and 
$F_i(\rho_1,\dots,\rho_n)$ and $ G_i(\rho_1,\dots,\rho_n)$ 
are the same as in Eq.(\ref{c9}).
The corresponding Fokker-Planck equation is ($G_{ij}=G_i \delta_{ij}$ where
$\delta_{ij}$ is the Kronecker delta-symbol)
\begin{eqnarray}\label{d10}
\frac{\partial}{\partial t} p = - \sum_{i=1} \frac{\partial}{\partial x_i}
\bigg \{ p \bigg[ F_i (x_1,\dots,x_n,t) + \nonumber \\
 \frac{1}{2}\sum_{j=1}^n
\sum_{k=1}^n G_{ij}(x_1,\dots,x_n,t)\frac{\partial }{\partial
x_i}[G_{ik}(x_1,\dots,x_n,t)] \bigg] \bigg \} + \nonumber \\ \frac{1}{2} \sum_{i=1}^n \sum_{j=1}^m
\frac{\partial}{\partial x_i} \frac{\partial}{\partial x_j}[p
G_{ij}(x_1,\dots,x_n,t)G_{ji}(x_1,\dots,x_n,t)]
\end{eqnarray}
We are interested in stationary solutions $p_s$ of Eq.(\ref{d10}). In the 
general case such solutions can be obtained only numerically. 
Analytic solutions can be obtained  when the conditions for detailed balance 
are satisfied \cite{gard1}. For the case of Eq.(\ref{d10}) the second of these
conditions is the same as the second condition from (\ref{c11}). $G_i$
for the case of Stratonovich is the same as $G_i$ from the case of Ito. Then
the second condition for existence of detailed balance is not satisfied. Then
one can hope to obtain approximate analytic solutions for particular
cases of Eq.(\ref{d10}) by the method of the adiabatic elimination discussed in
the previous section.
\par
Let us consider the following particular case of the Eqs.(\ref{b3}).
Let $\alpha_{ij}^0 = 0$ and $\alpha_{ijk}=0$. In addition let $\eta_2 =0$.
For the case of two populations we obtain the following system of equations
\begin{eqnarray}\label{d12}
d{\rho}_1 &=& [r_1^0 \rho_1 (1 + r_{11} \rho_1 + r_{12} \rho_2) + \rho_1/2] dt + \rho_1 dW_1
\nonumber \\
d{\rho}_2 &=& r_2^0 \rho_2 (1+ r_{21} \rho_1 + r_{22} \rho_2 )dt
\end{eqnarray}
Let again $\rho_2$ be the fast relaxing variable,i.e., $d \rho/dt$ tends to $0$
very fast in the time. Then in the second equation of Eqs.(\ref{d12})
one can set $d \rho /dt =0$ and then the resulting equation has solutions
$\rho_2 =0$ (extinction of the second population) or
\begin{equation}\label{d13}
\rho_2 = - \frac{1}{r_{22}} - \frac{r_{21}}{r_{22}}\rho_1
\end{equation}
which corresponds to a "slaving" of the "fast" variable $\rho_2$
by the "slow" variable $\rho_1$. The substitution of Eq.(\ref{d13})
in the first equation of Eqs.(\ref{d12}) leads to the stochastic differential
equation
\begin{equation}\label{d14}
d \rho_1 = \left[ r_1^0 \left( r_{11} - \frac{r_{12} r_{21}}{r_{22}} \right) 
\rho_1^2 +
\left[ r_1^0 \left( 1 - \frac{r_{12}}{r_{22}}\right) +1/2 \right] \rho_1 \right] dt + \rho_1 dW_1
\end{equation}
Eq.(\ref{d14}) can be treated by the methodology discussed above.
In order to obtain an analytic result we have to assume
\begin{equation}\label{d15}
r_1^0 = \frac{3/2}{1-r_{12}/r_{22}}
\end{equation}
The application of the methodology connected to \textit{Proposition 2}
leads to the distribution
\begin{equation}\label{d16}
p^0(\rho_1) = A \rho_1^2 \exp(2 \mu_2 \rho_1) 
\end{equation}
where
\begin{equation*}
A = \frac{\mu_2^3}{\left( \cfrac{\mu_2^2 b_2^2}{2}-\cfrac{\mu_2 b_2}{2}+
\cfrac{1}{4} \right) \exp(2 \mu_2 b_2) - \left( \cfrac{\mu_2^2 b_1^2}{2} -
\cfrac{\mu_2 b_1}{2} + \cfrac{1}{4} \right) \exp(2 \mu_2 b_1)} 
\end{equation*}
and
\begin{equation}\label{d17}
\mu_2 = \frac{3 r_{11} (r_{22}-r_{11})}{2(r_{22}-r_{12})} <0
\end{equation}

\section{Concluding remarks}
We note that the environment can influence not only the birth rates of the
interacting populations. The environment can influence also the interaction
coefficients. Thus the discussed above model is the simplest of the three
categories models of interacting populations: (i)
Models accounting for the influence of environment on the growth rates (one model
of this class is discussed in this paper); (ii)
Models accounting for the influence of the environment on the coefficients
of the interaction among the populations; and (iii)
Models accounting for the influence of the environment both on the growth rates
and the coefficients of interactions among the populations.
The equations for all classes of the models are discussed elsewhere \cite{bm1}.
\par
One result of our study above is that analytic solutions of the Fokker-Planck
equations connected to the dynamic of interacting populations can be obtained
for the case of one population. For two or more populations one can obtain
approximate solutions in some particular cases. In the general case one has to
solve the model equations numerically with the help of computers.
\begin{appendix}
\section{Nonlinear dynamics and interacting populations}
The nonlinear characteristics of the complex systems are intensively studied 
in different areas of science \cite{hak,feigenbaum} such for an example as the optics \cite{otsuka,
kang},
fluid mechanics \cite{vb}, biology \cite{volpert} or population dynamics
\cite{var}-\cite{je}, etc. \cite{pav} - \cite{d2}. Various mathematical methods connected to nonlinear time
series analysis \cite{ks} and nonlinear PDEs \cite{kudr05} - \cite{malf} are used in the study of these systems.
In this paper we discuss a class of models of the
the dynamics of interacting populations. These models consist of equations 
that contain only time dependence of the population densities.
What we add to the previous version of the models \cite{dv1}-\cite{dv5} is
an influence of the environment on the growth rates of the interacting
populations. This (random) influence has the following effect: 
instead of equations for the trajectories of the populations in the phase space of the population densities we shall 
write and solve equations for the probability density functions of the population 
densities. 
\section{Multiplicative white noise. Stochastic integrals of
Ito and Stratonovich kind}
Let us consider a system that is influenced by noise. The current state of the
system is $X(t)$ and the intensity of the noise depends on $X(t)$. 
Let the evolution of the system state be described by the stochastic 
differential equation
\begin{equation}\label{aa1}
{\dot{X}}(t) = f[X(t)] + \sigma[X(t)]  \zeta(t), \ X(0) = X_0
\end{equation}
If $\sigma[X(t)]=0$ then Eq.(\ref{aa1}) is deterministic one. If $\sigma[X(t)] = {\rm const}$
and if $\zeta(t)$ is Gaussian white noise then Eq.(\ref{aa1}) is equation of
Langevin kind. As a more general case $\sigma[X(t)]$ is not a constant
and if $\zeta(t)$ is a Gaussian white noise then Eq.(\ref{aa1}) describes the 
case of multiplicative Gaussian white noise. Below we shall discuss several
features of the solution of Eq.(\ref{aa1}) for the case of presence of the
multiplicative Gaussian white noise. 
\par
The formal integration of eq.(\ref{aa1}) leads to the integral equation  
\cite{wentzell,jetschke}
\begin{equation}\label{aa2}
X_t = X_0 + \int_0^t d \tau \ f(X_\tau) + \int_0^t dW_\tau \ \sigma(X_\tau),
\end{equation}
where $W_\tau$ is a Wiener process ($dW_\tau = \zeta(\tau) d \tau$).
The second integral from Eq.(\ref{aa2}), namely $\int_0^t dW_\tau \ \sigma(X_\tau)$,
is a stochastic integral. There are two interpretations of this integral:
(a) as integral of Ito kind; and (b) as integral of Stratonovich kind. It depends
on the characteristics of the modelled system which kind of integral has to be used.
\subsection{Interpretation of the stochastic integral as an integral of Ito kind}
Let us interpret the above stochastic integral as
\begin{equation}\label{aa3}
I_t = \int_0^t dW_\tau \sigma(X_\tau) = {\rm qa}\lim_{\delta_n \downarrow 0} I_t^{(n)}
\end{equation}
In Eq.(\ref{aa3}) qa $\lim_{\delta_n \downarrow 0}$ is a quadratic
average limit. This limit has to be understood as tendency to $0$ of the
expectation $E \mid I_t - I_t^{(n)} \mid^2$:
\begin{equation}\label{aa4}
\lim_{\delta_n \downarrow 0} E \mid I_t - I_t^{(n)}\mid^2 =0,
\end{equation}
where
\begin{equation}\label{aa5}
I_t^{(n)} = \sum_{i=0}^{n-1} \sigma(X_{t_{i}})(W_{t_{i+1}} - W_{t_{i}}),
\end{equation}
and $0 =t_0 < t_1 < \dots < t_n$; $\delta_n = \max_{i}(t_{i+1}-t_i)$. The
integral of kind (\ref{aa3}) is called integral of Ito kind. Then the
equation (\ref{aa2}) can be written in the following differential form
\begin{equation}\label{aa6}
dX_t = f(X_t)dt + \sigma(X_t) dW_t
\end{equation}
The initial condition is $X_0=X(0)$ and $X_0$ is a random variable which
probability density function is independent on the Wiener process $W_t$.
\subsection{Interpretation of the stochastic integral as an integral of
Stratonovich kind}
The model system can be of such kind that the stochastic integral
in Eq.(\ref{aa2}) is not an integral of Ito kind. This situation arises when 
for an example the noise process has a finite correlation time. Such processes
are present frequently in the real systems and the corresponding stochastic
integral is integral of Stratonovich kind. The interpretation of the
stochastic integral from Eq.(\ref{aa2}) for the last case is as follows:
\begin{equation}\label{aa7}
S_t = \int_0^t dW(\tau) \circ \sigma(X_\tau) = {\rm qa}{\lim_{\delta_n
\downarrow 0}
\sum_{i=0}^{n-1} \sigma \left[\frac{1}{2}(X_{t_{i+1}}+X_{t_{i}}) \right]}(W_{t_{i+1}}-
W_{t_{i}}).
\end{equation}
Then Eq.(\ref{aa2}) can be written in the following differential form
\begin{equation}\label{aa8}
dX_t = f(X_t) dt + \sigma(X_t) \circ dW_t.
\end{equation}
{\em Let $\sigma$ be continuous differentiable function}.
Then a relationship exists between the integrals of Ito and Stratonovich kind.
The relationship is as follows:
\begin{equation}\label{aa9}
\int_0^t dW_\tau \circ \sigma(X_\tau) = \int_0^t dW_\tau \sigma(X_\tau) +
\frac{1}{2} \int_0^t d\tau \sigma'(X_\tau) \sigma(X_\tau)
\end{equation}
We note that for the case of additive white noise $\sigma ={\rm const}$.
Then $\sigma' =0$ and the Ito integral coincides with the Stratonovich integral.
\par
We obtain on the basis of Eq.(\ref{aa9}) that the Stratonovich differential
equation (\ref{aa8}) is equivalent to the following stochastic differential equation of
Ito kind:
\begin{equation}\label{aa10}
dX_t = [f(X_t)+\frac{1}{2} \sigma'(X_t) \sigma(X_t)]dt + \sigma(X_t) dW_t
\end{equation}
\textit{In this paper we shall assume that $\sigma$ is continuously 
differentiable}.
\section{Probability density function for the case of multiplicative white
noise and Ito kind of stochastic differential equation}
In this case the solution $X_t$ of the Eq.(\ref{aa6}) is a Markov process and
the p.d.f. $p(x,t)$ for the values of $X$ (if the p.d.f. exists) is given by the
Fokker-Planck equation \cite{gard1}:
\begin{equation}\label{ab1}
\frac{\partial }{\partial t}p(x,t) = - \frac{\partial}{\partial x} [p(x,t) f(x)] +
\frac{1}{2} \frac{\partial^2}{\partial x^2}[p(x,t) \sigma^2(x)],
\end{equation} 
with initial condition $p(x,0) = p_0(x)$. Let us now discuss the behavior of the
solution $p(x,t)$ of Eq.(\ref{ab1}) at $t \to \infty$. We shall formulate an
Observation and for this we need the notion of natural boundary
point.
\par
Let the interval of possible values of the diffusion process $X$ that is
solution of Eq.(\ref{aa6}) be within the interval $[b_1,b_2]$. If $f$ and
$\sigma$ are continuously differentiable  in this interval
then the solution of Eq.(\ref{aa6}) exists till the time point when one of the
boundary points $b_{1,2}$ is reached. After that time the behavior of the
system depends on the boundary conditions. When the boundary point $b_1$ can't
be reached for finite time it is called inaccessible  (the
same is the situation with the point $b_2$). The inaccessible boundary point 
$b_1$ is called natural  when the solution $X(x_0)$ that starts from
$x_0 \in (b_1,c)$, $c<b_2$ accesses first the point $c$ with probability $1$.
This means that for $t \to \infty$ the point $b_1$ almost surely will be not
accessed.
\par
What follows is \cite{horsthemke}
\begin{flushleft}
{\bf Observation 1}:
\end{flushleft}
{\em
Let $b_1$ and $b_2$ be natural boundary points ($-\infty \le b_1 < b_2 \le 
\infty $). Let in addition $\sigma(x) >0$ in $(b_1,b_2)$. Then
the diffusion process $X_t$ that is solution of the stochastic differential
equation Eq.(\ref{aa6}) has unique invariant distribution with
p.d.f.
\begin{equation}\label{ba2}
p^0(x) = \frac{\cal{N}}{\sigma^2(x)} \exp \left( \int_c^x dy 
\frac{2 f(y)}{\sigma^2(y)} \right), \ \ \vee x \in (b_1,b_2)
\end{equation}
if the quantity
\begin{equation}\label{ba3}
{\cal{N}}^{-1}= \int_{b_1}^{b_2} dx \frac{1}{\sigma^2(x)} \exp \left( \int_c^x dy 
\frac{2 f(y)}{\sigma^2(y)} \right), \ \ b_1 < c < b_2
\end{equation}
has finite value. In addition each time-dependent solution $p(x,t)$ of the
Fokker-Planck equation (\ref{ab1}) in $(b_1,b_2)$ satisfies
\begin{equation}\label{ba4}
\lim_{t \to \infty} p(x,t) = p^0(x)
\end{equation}}
\par
Let us now consider the system of coupled stochastic 
equations
\begin{equation}\label{ba5}
{\dot{X}}_i(t) = f_i[X_1(t),\dots,X_n(t)] + \sum_{j=1}^m g_{ij}[X_1(t),
\dots,X_n(t)] \zeta_j(t), \ i=1,\dots,n
\end{equation}
where $\zeta_j(t)$ are independent white Gaussian noises. If the arising
in the process of solution of Eq.(\ref{ba5}) stochastic integrals are of Ito
kind then one has to solve the system of coupled stochastic differential
equations
\begin{equation}\label{ba6}
dX_i(t) = f_i[X_1(t),\dots,X_n(t)] + \sum_{j=1}^m g_{ij}[X_1(t),
\dots,X_n(t)] dW_j(t), \ i=1,\dots,n
\end{equation}
where $W_j(t)$ are independent Wiener processes.  For the conditional
probability density $p=p(x_1,\dots,x_n,t \mid {x_0}_1, \dots, {x_0}_n,t)$ of 
${\bf X}=(X_1,\dots,X_n)$ one obtains the Fokker-Planck equation
\begin{eqnarray}\label{ba7}
\frac{\partial}{\partial t} p = - \sum_{i=1} \frac{\partial}{\partial x_i}
[p f_i (x_1,\dots,x_n,t)] + \nonumber \\ \frac{1}{2} \sum_{i=1}^n \sum_{j=1}^m
\frac{\partial}{\partial x_i} \frac{\partial}{\partial x_j}[p
g_{ij}(x_1,\dots,x_n,t)g_{ji}(x_1,\dots,x_n,t)]
\end{eqnarray}
\section{Probability density function for the case of multiplicative white noise
and Stratonovich kind of stochastic differential equation}
In this case the solution $X_t$ of the Eq.(\ref{aa10}) is a Markov process and
the p.d.f. $p(x,t)$ for the values of $X$ (if the p.d.f. exists) is given by the
Fokker-Planck equation \cite{wentzell,jetschke}:
\begin{equation}\label{ac1}
\frac{\partial }{\partial t}p(x,t) = - \frac{\partial}{\partial x} [p(x,t) f^*(x)] +
\frac{1}{2} \frac{\partial^2}{\partial x^2}[p(x,t) \sigma^2(x)],
\end{equation} 
where 
\begin{equation}\label{ac2}
f^*(X_t)=f(X_t)+\frac{1}{2} \sigma'(X_t) \sigma(X_t)
\end{equation}
with initial condition $p(x,0) = p_0(x)$. The behavior of the
solution $p(x,t)$ of Eq.(\ref{ab1}) at $t \to \infty$ is as follows. 
Let the interval of possible values of the diffusion process $X$ that is
solution of Eq.(\ref{aa6}) be within the interval $[b_1,b_2]$. 
Let $f^*$ and $\sigma$ are continuously differentiable 
 in this interval
then the solution of Eq(\ref{aa6}) exists till the time point when one of the
boundary points $b_{1,2}$ is accessed. Then we can formulate \cite{horsthemke}
\begin{flushleft}
{\bf Observation 2}:
\end{flushleft}
{\em
Let $b_1$ and $b_2$ be natural boundary points ($-\infty \le b_1 < b_2 \le 
\infty $). Let in addition $\sigma(x) >0$ in $(b_1,b_2)$. Then
the diffusion process $X_t$ that is solution of the stochastic differential
equation Eq.(\ref{aa6}) has unique invariant distribution with
p.d.f.
\begin{equation}\label{ca2}
p^0(x) = \frac{\cal{N}}{\sigma^2(x)} \exp \left( \int_c^x dy 
\frac{2 f^*(y)}{\sigma^2(y)} \right), \ \ \vee x \in (b_1,b_2)
\end{equation}
if the quantity
\begin{equation}\label{ca3}
{\cal{N}}^{-1}= \int_{b_1}^{b_2} dx \frac{1}{\sigma^2(x)} \exp \left( \int_c^x dy 
\frac{2 f^*(y)}{\sigma^2(y)} \right), \ \ b_1 < c < b_2
\end{equation}
has finite value. In addition each time-dependent solution $p(x,t)$ of the
Fokker-Planck equation (\ref{ac1}) in $(b_1,b_2)$ satisfies
\begin{equation}\label{ca4}
\lim_{t \to \infty} p(x,t) = p^0(x)
\end{equation}}
\par
Let us now consider the system of coupled stochastic 
equations
\begin{equation}\label{ca5}
{\dot{X}}_i(t) = f_i[X_1(t),\dots,X_n(t)] + \sum_{j=1}^m g_{ij}[X_1(t),
\dots,X_n(t)] \zeta_j(t), \ i=1,\dots,n
\end{equation}
where $\zeta_j(t)$ are independent white Gaussian noises. If the arising
in the process of solution of Eq.(\ref{ba5}) stochastic integrals are of
Stratonovich
kind then one has to solve the system of coupled stochastic differential
equations
\begin{eqnarray}\label{ca6}
dX_i(t) = \bigg \{ f_i[X_1(t),\dots,X_n(t)] + \nonumber \\
\frac{1}{2}\sum_{j=1}^n
\sum_{k=1}^n g_{ij}(X_1(t),\dots,X_n(t))\frac{\partial }{\partial
x_i}[g_{ik}(X_1(t),\dots,X_n(t))] \bigg \} 
+ \nonumber \\ \sum_{j=1}^m g_{ij}[X_1(t),
\dots,X_n(t)] dW_j(t), \ i=1,\dots,n
\end{eqnarray}
where $W_j(t)$ are independent Wiener processes.  For the conditional
probability density $p=p(x_1,\dots,x_n,t \mid {x_0}_1, \dots, {x_0}_n,t)$ of 
${\bf X}=(X_1,\dots,X_n)$ one has the Fokker-Planck equation
\begin{eqnarray}\label{ca7}
\frac{\partial}{\partial t} p = - \sum_{i=1} \frac{\partial}{\partial x_i}
\bigg \{ p \bigg[ f_i (x_1,\dots,x_n,t) + \nonumber \\
 \frac{1}{2}\sum_{j=1}^n
\sum_{k=1}^n g_{jk}(x_1,\dots,x_n,t)\frac{\partial }{\partial
x_j}[g_{jk}(x_1,\dots,x_n,t)] \bigg] \bigg \} + \nonumber \\ \frac{1}{2} \sum_{i=1}^n \sum_{j=1}^m
\frac{\partial}{\partial x_i} \frac{\partial}{\partial x_j}[p
g_{ij}(x_1,\dots,x_n,t)g_{ji}(x_1,\dots,x_n,t)]
\end{eqnarray}
\end{appendix}

\end{document}